\documentclass[prl, aps, twocolumn, superscriptaddress, 10pt]{revtex4-1}

\usepackage[utf8]{inputenc} 
\usepackage[T1]{fontenc} 
\usepackage[english]{babel} 
\usepackage{lmodern} 
\usepackage{geometry} 
\usepackage{setspace} 
\usepackage{indentfirst} 
\usepackage{graphicx} 
\usepackage{hyperref} 
\hypersetup{
	colorlinks=true,
	linkcolor=blue,
	citecolor=blue,
	urlcolor=blue
}
\usepackage{natbib} 
\usepackage{verbatim} 
\usepackage[pangram]{blindtext} 

\usepackage{amsmath, amsthm, amscd, amssymb} 
\usepackage[all,cmtip]{xy} 
\usepackage{listings} 
\usepackage{physics} 
\usepackage{lipsum}

\def \dif {\mathrm{d}}

\makeindex

\begin{document}
\title{Extending vacuum trapping to absorbing objects with hybrid\\
Paul-optical traps}

\author{Gerard P. Conangla}\email{gerard.planes@icfo.eu}
\affiliation{ICFO Institut de Ciències Fotòniques, Mediterranean Technology Park, 08860 Castelldefels (Barcelona), Spain}

\author{Raúl A. Rica}\email{rul@ugr.es}
\affiliation{ICFO Institut de Ciències Fotòniques, Mediterranean Technology Park, 08860 Castelldefels (Barcelona), Spain}
\affiliation{Universidad de Granada, Department of Applied Physics, Faculty of Sciences, 18071, Granada, Spain}

\author{Romain Quidant}\email{romain.quidant@icfo.eu}
\affiliation{ICFO Institut de Ciències Fotòniques, Mediterranean Technology Park, 08860 Castelldefels (Barcelona), Spain}
\affiliation{ICREA-Institució Catalana de Recerca i Estudis Avançats, 08010 Barcelona, Spain}
\date{\today}

\begin{abstract}
The levitation of condensed matter in vacuum allows the study of its physical properties under extreme isolation from the environment. It also offers a venue to investigate quantum mechanics with large systems, at the transition between the quantum and classical worlds. In this work, we study a novel hybrid levitation platform that combines a Paul trap with a weak but highly focused laser beam, a configuration that integrates a deep potential with excellent confinement and motion detection. 
We combine simulations and experiments to demonstrate the potential of this approach to extend vacuum trapping and interrogation to a broader range of nanomaterials, such as absorbing particles. We study the stability and dynamics of different specimens, like fluorescent dielectric crystals and gold nanorods, and demonstrate stable trapping down to pressures of 1 mbar.
\end{abstract}

\maketitle

\section{Introduction} 
The study of micro- and nano-sized systems often requires good isolation from the environment. This can be challenging, because they usually either lie on a substrate or are surrounded by a liquid~\citep{muskens2007strong}, which can significantly alter their intrinsic physical and chemical properties. Moreover, their small dimensions are often associated with weak signals, difficult to separate from environmental noise. A possible solution is to hold the specimen of interest in a trap, preferentially at low pressures~\citep{ashkin1976optical, benabid2002particle, santesson2004airborne, kane2010levitated}.

Particle traps can be realized in a number of ways. Paul traps use a combination of AC and DC electric fields for the confinement of charged particles in air and vacuum~\citep{Paul1990}. Their use for the study of micro and nanoparticles is nowadays widespread~\citep{wuerker1959electrodynamic, schlemmer2004interaction,grimm2006charging,kane2010levitated,bell2014single,howder2015thermally, conangla2018motion}, for instance to investigate the optical properties of atmospheric aerosol droplets~\citep{krieger2012exploring, davies2012time}. Optical tweezers---or, more generally, optical dipole traps---allow one to trap dielectric particles near the maximum of a light intensity distribution, like the focus of a strongly focused laser beam~\citep{ashkin1970acceleration, ashkin1986observation,grier2003revolution}. Although mostly used to trap dielectric microparticles suspended in a liquid~\citep{svoboda1994biological}, optical tweezers can also levitate micro and nanoparticles in air or vacuum~\citep{Li2010,Gieseler2012}.

Nevertheless, both approaches have limitations that restrict their use to specific nanoparticle types and interrogation schemes. Optical tweezers, on the one hand, require high powers to trap in vacuum, typically $\sim 100$~mW~\citep{Gieseler2012, millen2019optomechanics}. Such powers are responsible for substantial heating~\citep{millen2014nanoscale}, leading to particle photodamage already at pressures of a few tens of millibars~\citep{Neukirch2015}. Hence, low damping regimes---often the most interesting for fundamental studies---can only be accessed with low absorbing materials like silica~\citep{Jain2016}. On the other hand, Paul traps have low spatial confinement~\citep{alda2016trapping,bykov2019direct}, hence limiting the ability to interact with the trapped specimen and detect its position accurately. 

\begin{figure}
\begin{center}
\includegraphics[width=0.46\textwidth]{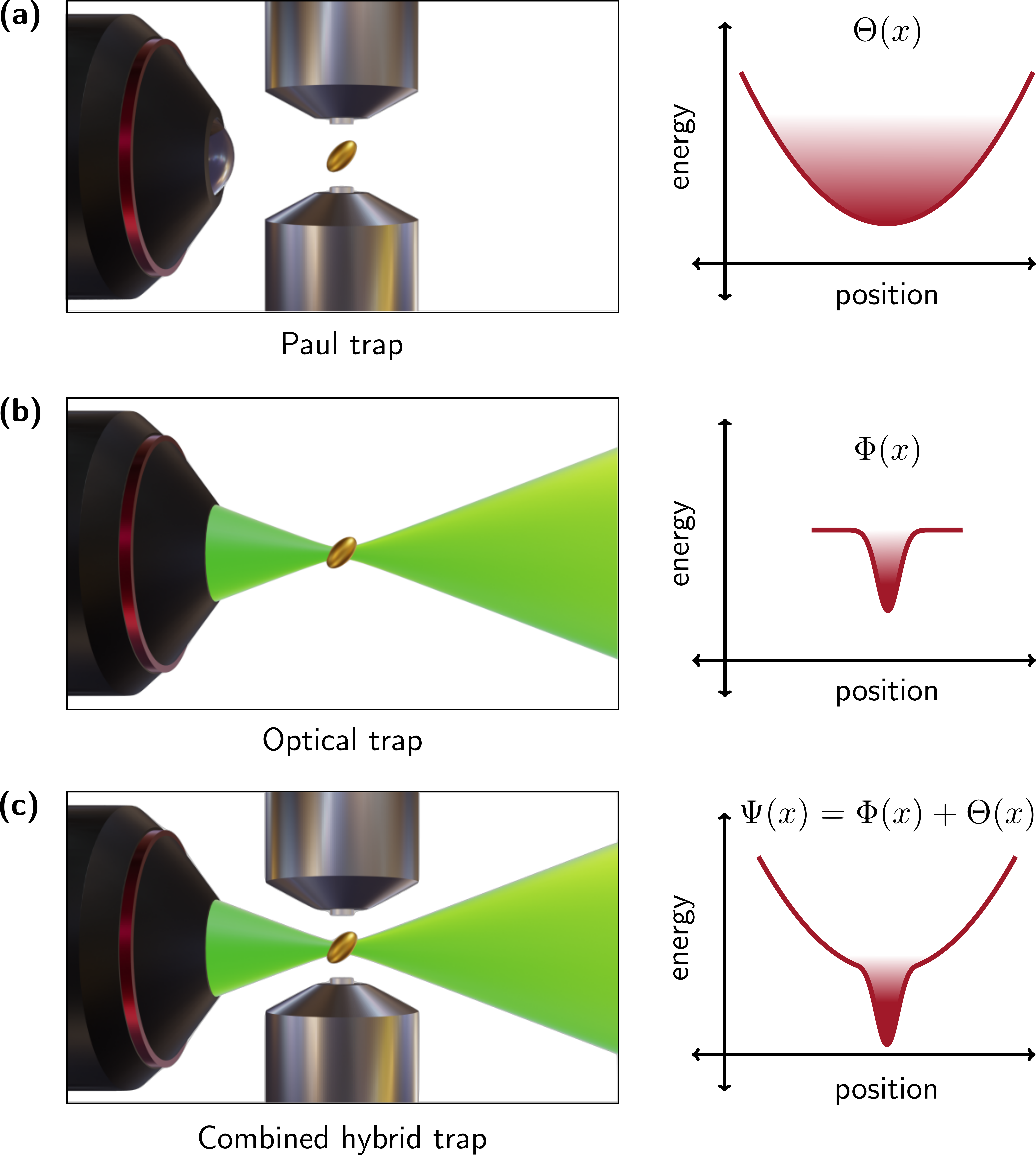}
\caption[Hybrid trap]{\textbf{Sketch of the hybrid Paul-optical trap}. 
(a) A charged nanoparticle (here a gold nanorod) is trapped in a deep quadratic potential $\Theta(x)$, created by the Paul trap. The particle may explore a large volume because the potential has a small gradient. (b) The particle is now optically trapped in potential $\Phi(x)$. The particle is tightly confined, but the potential barrier is low. (c) Activating both the Paul trap and the optical trap, the particle is stored in a dimple potential $\Psi(x) = \Theta(x) + \Phi(x)$.}
\label{fig:1}
\end{center}
\end{figure}

Hybrid traps present a possible workaround beyond Paul traps and optical tweezers, which adds further flexibility by combining two types of fields. For instance, trapping of micro- and nanoparticles was reported in magneto-gravitational traps~\citep{hsu2016cooling, slezak2018cooling, houlton2018axisymmetric}, as well as in a Paul trap interfaced to a Fabry-Perot cavity~\citep{millen2015cavity, fonseca2016nonlinear}. 

Here, we implement and characterize a hybrid platform combining Paul and optical traps (see Fig.~\ref{fig:1}). By superimposing a weak but tightly focused laser beam to the potential created by a Paul trap, we form a \emph{dimple} trap~\citep{weber2003bose}, combining high particle confinement with reduced bulk heating. We demonstrate that our platform is versatile both for levitating and optically interrogating particles with high optical absorption. In particular, we validate the system with gold nanoparticles, nanodiamonds and crystals hosting rare earth ions. We also study numerically and experimentally the particle dynamics in terms of its position distribution at equilibrium, which we use to reconstruct the trapping potential.

\section{Experimental setup}

The hybrid Paul-optical platform is sketched in Fig.~\ref{fig:1}~(c). It combines an end-cap Paul trap made of two steel electrodes with rotational symmetry, with a 0.8 NA objective lens focusing a 532 nm laser beam (power $P \leq 20$ mW). The electrodes were designed to provide high optical access and a linear electric field in a large volume around the trapping region. Typical parameter ranges for the frequency and amplitude of the driving AC field were $1$~kHz -- $30$~kHz and $0.6$ $\text{kV}_\text{pp}$ -- $2$ $\text{kV}_\text{pp}$, respectively. A piezoelectric stage was used to adjust the position of the electrodes with respect to the laser focus with sub-micrometer accuracy. The trap, the focusing objective and a lens used to collect the forward scattered light are mounted inside a vacuum chamber. 

All particle types were prepared in ethanol suspensions to load them into the trap. Loading was realized at ambient pressure with a custom-made electrospray, illuminating the trapping volume with a weakly focused 980 nm laser to detect incoming particles. Once a single one was trapped in the Paul trap, we turned on the 532 nm laser , responsible for the optical potential of the hybrid trap. Together, the Paul trap's electric field and the optical gradient force from the focused laser beam produce a dimple-like effective potential, as illustrated in Fig.~\ref{fig:1}~(c). In this situation, the particle is confined to a region that is significantly smaller than what would be achieved with only the Paul trap. Notice that the laser beam alone would not be able to keep the particle trapped at these low powers.

\begin{figure}
\begin{center}
\includegraphics[width=0.47\textwidth]{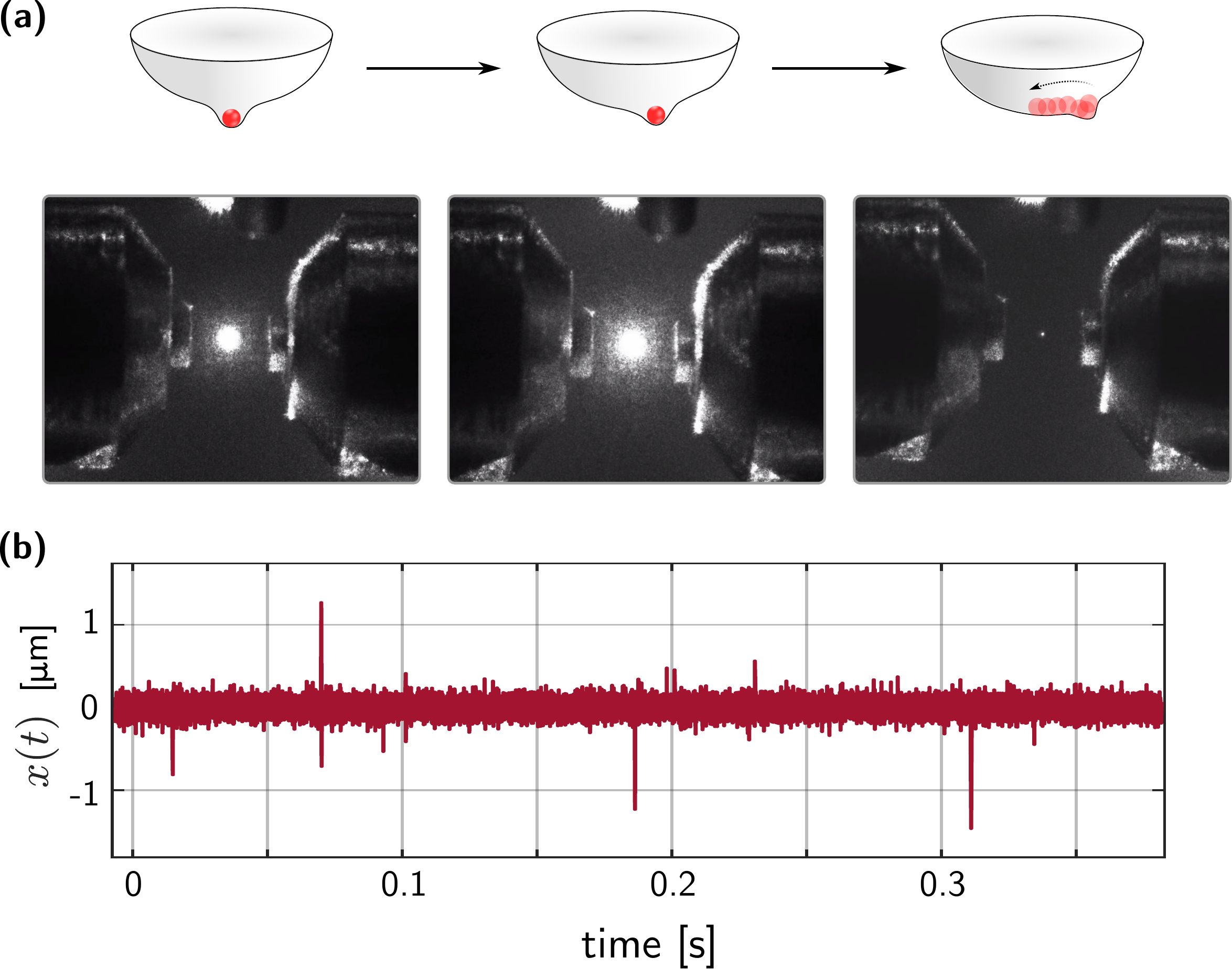}
\caption[Potential displacement]{\textbf{Changing the relative position between Paul and optical traps}.
(a) The Paul trap is displaced with respect to the optical field with a piezoelectric nanopositioner. If this displacement is small, the trapped particle stays in the potential minimum (i.e., at the focus of the beam). However, if the displacement is increased further, the Paul trap eventually pushes the particle back to its center and its brightness decreases suddenly. The photographs portray the Paul trap moving towards the incoming beam. In this case, the maximum of the particle brightness takes place when the Paul trap and the optical trap are slightly misaligned, and the Paul trap exactly compensates the optical scattering force.
(b) Time trace of a particle in the hybrid trap (upper-left picture situation) at low optical power (1 mW of 532 nm light). The spikes indicate that the particle hops in and out of the focus.
}
\label{fig:2}
\end{center}
\end{figure}

The relative position between the Paul and optical traps must be accurately adjusted to ensure that the potential minima are very close to each other. Otherwise, the particle crosses the optical potential barrier (much shallower than the Paul trap potential) and its dynamics are solely governed by the Paul trap electric field, as shown in Fig.~\ref{fig:2}. 

Position detection is achieved by interferometric measurements of the particle scattered light. When the particle is located in the optical dimple, the 532~nm forward-scattered light is collected and directed to a quadrant photodiode, which returns electric signals that are proportional to the particle motion in the 3 perpendicular directions $x(t)$ (parallel to trap axis), $y(t)$ (gravity direction) and $z(t)$ (beam propagation). 

\section{Results and analysis}

\begin{figure}
\begin{center}
\includegraphics[width=0.40\textwidth]{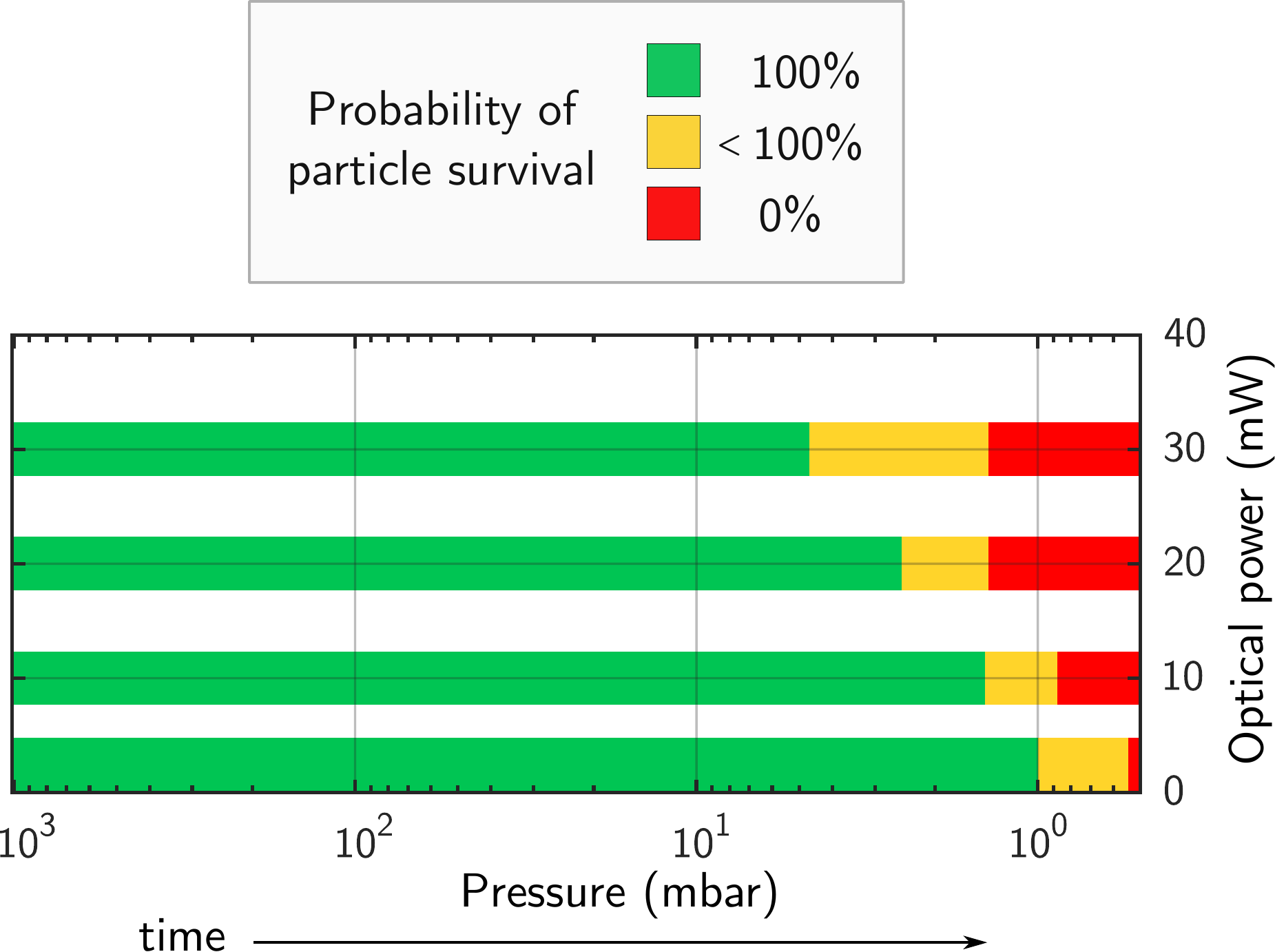}
\caption[Gold nanoparticles]{\textbf{Probability of survival of gold nanorods}.
Experimental probability of survival of a levitated gold nanorod at the focus of the hybrid trap when the pressure is progressively reduced (using 532 nm light). Green color marks that all particles achieved said pressure, yellow marks the survival of some (but not all) of the trapped particles, and red that no particle could be brought to that pressure. Most of the gold nanoparticles are lost around 1 mbar. We repeated the experiment with 1064 nm light, obtaining similar results.
}
\label{fig:3}
\end{center}
\end{figure}

We tested the hybrid trap with both dielectric and metallic particles. Gold nanorods of $33\text{ nm}\times63\text{ nm}$ were loaded into the Paul trap at ambient pressure, illuminated with the 532~nm beam, and later on brought to vacuum to assess its survival and stability. In this situation, the particle cannot dissipate efficiently all the radiation it absorbs from the laser, and its bulk temperature increases considerably. As can be seen in Fig. \ref{fig:3}, the gold particles could be maintained in the trap down to 10 mbar. Below this pressure, and depending on the power of the optical trap, the probability of the nanorods disappearing from the trap increased dramatically. At $\sim 1$ mbar most of the studied particles vanished, except for low optical powers (below $\sim 3$ mW). 
In the latter case, it was not possible to maintain a stably trapped particle: it hopped in and out of the focus intermittently and thus received a considerably lower average optical power. 
We also observed that, when the particles disappeared from the trap, the event happened rather suddenly, without clear prior changes in particle brightness. We speculate that this is due to the evaporation of the outer layer of the gold particle, which leads to a sudden change in charge-to-mass ratio and to unstable trapping conditions~\citep{Paul1990}.

To study the dynamics of dielectric particles in the hybrid trap, we analyze the dominant forces in the absence of absorption effects. The center of mass (COM) Langevin equation of motion of a particle in an electric quadrupole-optical trap can be obtained with Newton's 2nd law:
\begin{align}\label{eq:motion}
m\ddot{x} + m\Gamma \dot{x} - \frac{QV}{d^2}\cos \omega_\text{d} t\cdot x + \alpha\frac{1}{2}\nabla |\mathbf{E}|^2 = \sigma \eta(t).
\end{align} 
Here, $x(t)$ is the COM motion, $m$ is the particle mass~\citep{ricci2019accurate} and $\Gamma$ is the damping rate due to the interaction with residual gas molecules. From the Paul trap, $\omega_\text{d}$ is the trap driving frequency, $Q$ is the particle charge, $V$ the trap voltage amplitude and $d$ is the characteristic size. Regarding the optical trap, $\alpha$ is the particle polarizability and $\mathbf{E}\triangleq \mathbf{E}(x,y,z)$ is the optical field, which we approximate as a focused Gaussian Beam (see \hyperref[sec:supplemental]{Supplemental material}) to perform numerical simulations. Finally, $\sigma \eta(t)$ is a stochastic force that accounts for the thermal coupling with the environment, with $\eta(t)$ being a unit intensity Gaussian white noise and $\sigma = \sqrt{2 k_\text{B} T m\Gamma}$~\citep{Kubo1966,conangla2020overdamped}. 

In Eq.~\eqref{eq:motion} we do not consider gravity, which is small compared to the electric force (from the Paul trap) and the gradient force (from the optical tweezers field), nor the scattering force, that can be neglected for nanoparticles in tightly focused beams---as is our case when the particle is in a stable state in the optical trap~\footnote{When the particle is \emph{pushed} away from the center by the Paul trap, as shown in Fig.~\ref{fig:1}, the scattering force might become apparent.}. Since both relevant deterministic forces are gradients, the total potential will be the addition of the two individual trap potentials. However, notice that at low laser powers the levitated particle explores large regions of the optical trap and the dipole force cannot be approximated as a linear restoring force, as is common practice in optical tweezers experiments (i.e., there will be significant nonlinearities in the dynamics). The Paul trap potential, being much larger, can still be safely approximated as a quadratic (time-varying) potential. 

\begin{figure}
\begin{center}
\includegraphics[width=0.48\textwidth]{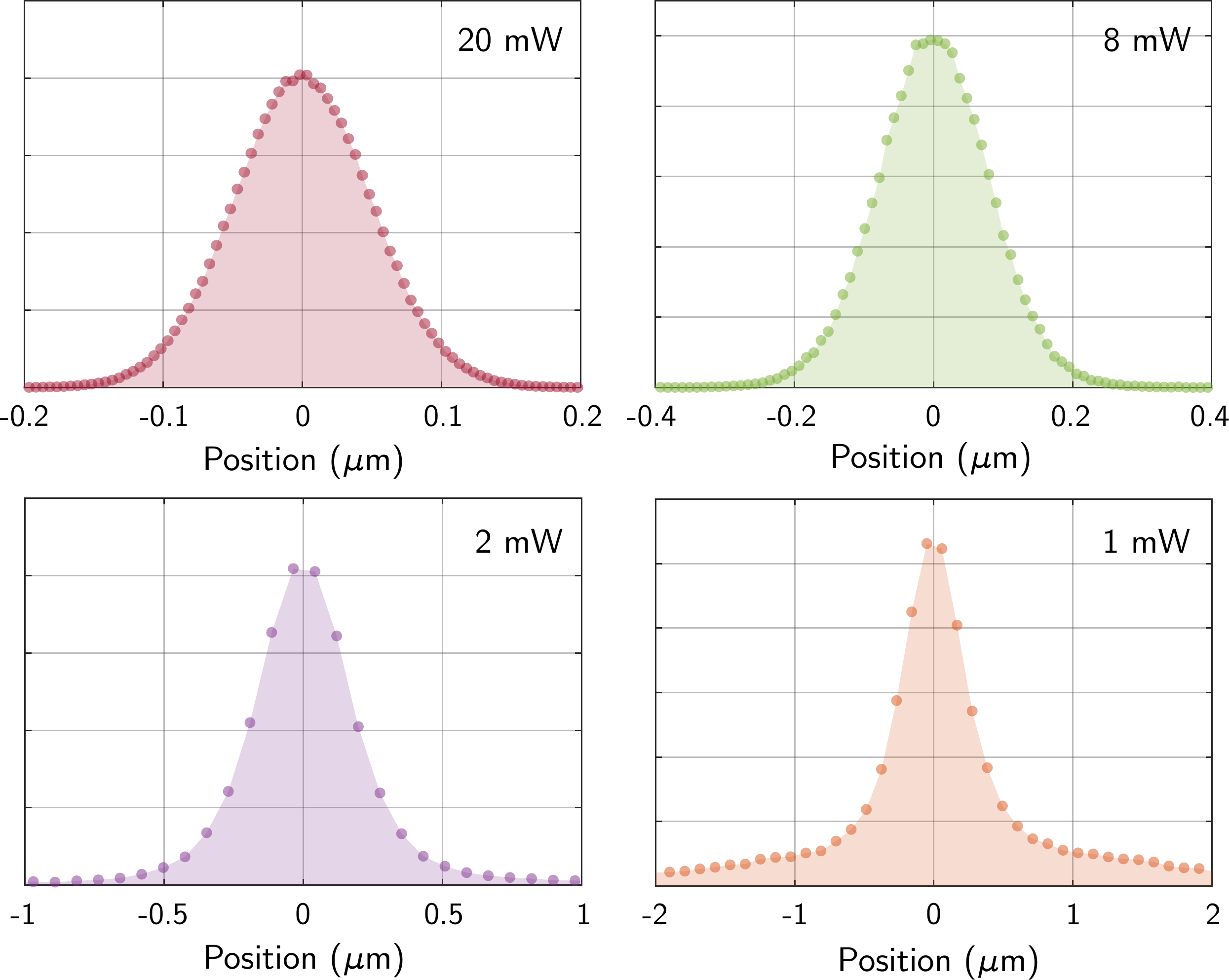}
\caption[Potential simulations]{\textbf{Surge of nonlinearities at low intensities}. 
The plots show numerical simulations of the position density distribution $\rho_\infty(x)$ of 100 nm dielectric nanoparticles (no absorption considered) for decreasing optical powers in the hybrid trap ($\lambda = 1064$ nm). Below 20 mW, a reduction in the laser power brings about position distributions that deviate more and more from a normal distribution (quadratic potential case, exemplified with the upper left plot). At 1 mW, the particle leaves and re-enters the beam focus intermittently, resulting in long tails in $\rho_\infty(x)$. Notice that the $x$-axis scale varies along the different plots.
}
\label{fig:4}
\end{center}
\end{figure}

\begin{figure*}
\begin{center}
\includegraphics[width=0.9\textwidth]{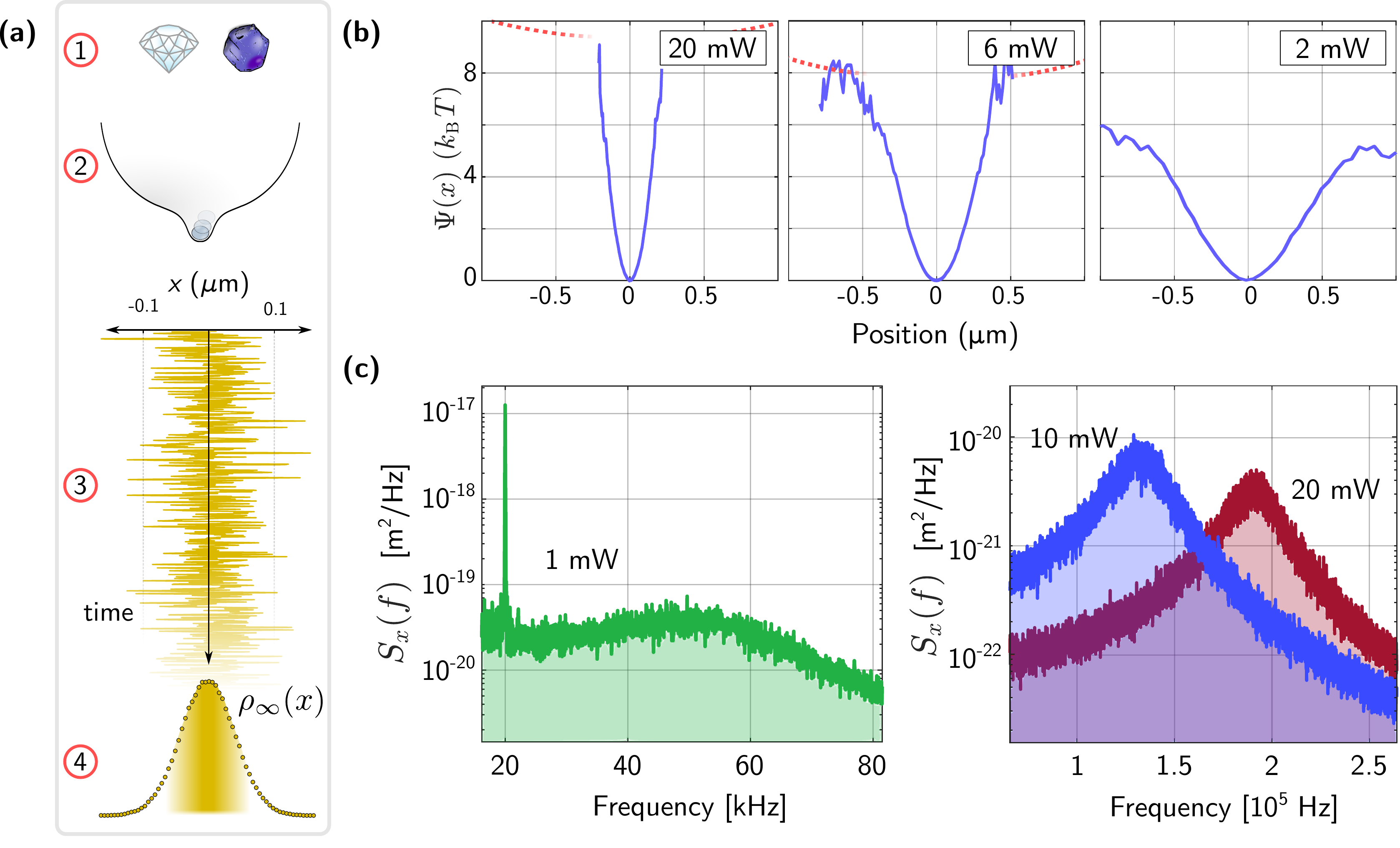}
\caption[Potential reconstruction]{\textbf{Potential reconstruction}. 
(a) Process of potential reconstruction. A dielectric particle (1) oscillates in the hybrid potential (2). A time trace $x(t)$ of the trapped particle is measured for a few seconds (3). From the time trace, we compute an histogram of the position (4) to estimate the Gibbs probability density function $\rho_{\infty} (x)$ (see Eq.~\eqref{eq:Gibbs}). From this density function, we may obtain $\Psi(x) = \frac{-1}{\beta}\ln\left(Z \rho_{\infty} (x)\right)$
(b) Reconstructed trap potentials for three different powers of 532 nm laser light (blue line) with a levitated nanodiamond ($r \sim 70$ nm). The dashed red lines are only a guide to the eye for the Paul trap section of the combined potential.
(c) PSDs of the motion at different laser powers (pressure inside the vacuum chamber of 50 mbar, $\lambda = 532$ nm). At 1 mW the optical field is not strong enough to keep the particle in the dimple, and the driving ($\omega_\text{d} = 20$ kHz) dominates the dynamics. At higher laser powers, the particle is trapped in the dimple of the potential well, where it oscillates driven by Brownian motion. As expected, both the trap frequency and the confinement increase with laser power.
}
\label{fig:5}
\end{center}
\end{figure*}

If we consider the adiabatic approximation for the Paul trap, i.e. we approximate the time-varying potential by an effective constant potential~\citep{conangla2020overdamped}, then at equilibrium $x(t)$ will follow the Gibbs probability density function
\begin{align}\label{eq:Gibbs}
\rho_{\infty} (x) = \frac{1}{Z} \exp ( - \beta \Psi (x)),
\end{align}
where $Z$ is a normalizing factor, $\beta = 1/k_\text{B}T$, and $\Psi(x)$ is the combined hybrid potential, introduced in Fig.~\ref{fig:1}. Hence, by estimating $\rho_{\infty} (x)$ and inverting Eq.~\eqref{eq:Gibbs} we can recover the effective potential of the hybrid trap.

This process is shown in Fig.~\ref{fig:4} with simulated time traces. Here, we studied theoretically the effects of a decrease in the optical field's intensity. The simulations reveal that when the laser power is below 8 mW, the position histograms start to deviate noticeably from a normal distribution. This is due to a reduced particle confinement, that lets the particle explore the nonlinearities away from the optical trap center. Eventually, at sufficiently low powers, the particle starts leaving and re-entering the beam focus. 

We observed a similar behaviour experimentally. Starting at a high power (e.g., 20 mW of 532 nm laser light), a progressive reduction of the power lead to a blinking of the particle, which we interpret as the particle hopping in and out of the optical trap. Experimentally, the blinking started at higher powers than in the simulations, but this was expected because the two potentials (optical and electric) are never perfectly aligned. Hence, the driving from the Paul trap may push the particle away from the focus before than in the idealized simulation.

The potential reconstruction from experimentally trapped dielectric particles is shown in Fig.~\ref{fig:5}, plotting the profiles obtained at different powers. The potential has the shape of a dimple trap, consisting of a large, flatter trapping volume (created by the Paul trap), superimposed to a tighter optical dipole trap, as represented in Fig.~\ref{fig:5}~(b) with data from a levitated nanodiamond. In these time traces, we filtered out the contribution of the driving AC field, which is never completely eliminated through field compensation and perturbs the results. The presence of non-linearities in the detection system (i.e., in the correspondence between detected signal and position) also posed a problem for such potential reconstructions. To correct for this, we eliminated fractions of the recorded time traces in which it was clear that the particle was away from the trap center, and inverted the remaining data with the expression found in Gittes et al.~\citep{gittes1998interference}. 

PSDs of the particle motion are also shown in Fig.~\ref{fig:5}~(c) for different optical powers. At 1 mW, the optical field is too weak to keep the particle in the focus. As a result, the particle hops in and out of the dimple and the dynamics are dominated by the Paul trap driving (peak at $\omega_\textbf{d} = 20$ kHz). This hopping disappears as the laser power is increased to higher values, and is almost absent for powers as low as 10 mW. In this situation, the particle oscillates with a frequency in the range of hundreds of kHz, significantly larger than in typical experiments with nanoparticles in Paul traps~\citep{conangla2018motion,bykov2019direct}. 

\section{Conclusions}
In conclusion, we built a hybrid nanoparticle trap by combining a Paul trap with a weak but highly focused optical beam, and demonstrated its suitability as an experimental platform to store highly absorbing particle species~\citep{hansen2005expanding, Rahman2017, conangla2018motion}. Even though our experiments were performed with nanoparticles, similar results are expected with larger particles of up to a few microns~\citep{krieger2012exploring,gong2018optical}. 

We validated our platform with gold, diamond and Erbium-doped nanoparticles, trapping and detecting them at optical powers below 10 mW, which are low compared to typical optical tweezers powers ($\sim100$ mW). We also verified that the hybrid scheme easily allows a reduction of the vacuum pressure, even below $\sim 5$ mbar with gold nanoparticles. This contrasts with previous optical levitation experiments in vacuum, where due to optical absorption of the trapped particles~\citep{millen2014nanoscale, hebestreit2018measuring}, the range of materials was limited to just a few options. In our hybrid scheme there is still room to reduce the heating, since in most experiments the presence of the optical field is only required to perform short measurements. To store the particles the Paul trap suffices, allowing for arbitrarily long levitation times~\citep{conangla2018motion}. 

Moreover, we used the measured position traces to reconstruct the effective potential with the Boltzmann--Gibbs distribution, comparing the results to numerical simulations of the stochastic equation of motion. As is intuitively expected, the trap potential has a dip in the middle, corresponding to the optical dipole gradient potential.

Due to its versatility, the studied hybrid platform is suited to investigate small objects under experimental constraints that are difficult to overcome with other trapping techniques. In aerosol science, hybrid traps could improve the confinement of the levitated droplets, boosting the sensitivity of the Raman and infrared spectra signals that are commonly targeted~\citep{krieger2012exploring}. 
Another promising direction is the study of isolated particles that strongly interact with light, such as metals~\citep{Schell2017flying} or particles with internal degrees of freedom~\citep{Rahman2017,conangla2018motion}. Prime examples of these are diamonds with color centers, quantum dots or crystals hosting rare earth ions, which live in a domain that is mostly classical but, still, not completely free of quantum effects~\citep{delord2020spin}. Finally, the ability to both move the optical trap and change its depth, in combination with control over the Paul trap, could be used to experimentally explore complex dynamics in bistable potentials~\citep{ricci2017optically,rondin2017direct,torres2020theoretical}

\vspace{0.5cm}
\textbf{Acknowledgements}.\hspace{0.2cm}The authors acknowledge Luís Carlos and Mengistie Debasu from the University of Aveiro for samples of the Er3+-doped particles. The authors acknowledge financial support from the European Research Council through grant QnanoMECA (CoG - 64790), Fundació Privada Cellex, CERCA Programme / Generalitat de Catalunya and the Severo Ochoa Programme for Centres of Excellence in R$\&$D (SEV-2015-0522), grant FIS2016-80293-R. R.A.R. also acknowledges financial support from the Junta de Andalucía for the project P18-FR-3583, the Spanish Ministry of Economy and Competitiveness for the projects  IJCI-2015-26091 and PGC2018-098770-B-I00, and the University of Granada for the project PPJI2018.12.

\bibliographystyle{apsrev4-1}
\bibliography{hybrid_trap}

\cleardoublepage
\section{Supplemental material}\label{sec:supplemental}

\phantomsection
\subsection{Experimental methods}\label{sec:setup}

\subsubsection{Electrospray}
The electrospray ensures that particles are highly charged ($50 < n < 1000$ of net e$^+$ charges in this study, depending on the particle). To detect the presence of trapped particles during the particle loading process, we used a weakly focused 980 nm diode laser to illuminate the trapping volume (focus spot $> 10 \, \mu$m).

\subsubsection{Position signal processing}
After the $x(t)$, $y(t)$ and $z(t)$ signals are acquired in the quadrant photodetector, they are sent to an FPGA for pre-processing and finally recorded in a computer. Additionally, if the particle has internal transitions or interacts strongly with the 532~nm laser light (e.g. highly absorbing materials like our gold nanorods), the fluorescence can also be collected and sent to a light-tight box. There, we can measure it with a fluorescence camera or a spectrometer.

\subsubsection{Compensation of stray fields}
Stray DC fields may push a levitated particle slightly away from the center of the Paul trap. The most common protocol to cancel these fields consists in creating an opposite DC voltage with the compensation electrodes, by checking that excess micromotion---a forced oscillation at the trap driving frequency that can be seen as a peak in the motion's power spectral density (PSD)---is minimized~\citep{berkeland1998minimization}. However, by adding a DC electric field with the electrodes, a particle at the center of the optical trap will also move away from the focus. In other words, cancelling stray DC fields \emph{and} pushing the particle away from the focus have the same effect in detection, and thus are difficult to distinguish. To decouple these two effects, we detect and move the particle with the same system: a $4f$ pair of lenses. Once the particle is trapped with the 532~nm laser, the $4f$ system can adjust the angle at which the laser beam enters the back of the trapping objective, which effectively changes the position of the beam focus in the focal plane.

\begin{figure}[h]
\begin{center}
\includegraphics[width=0.46\textwidth]{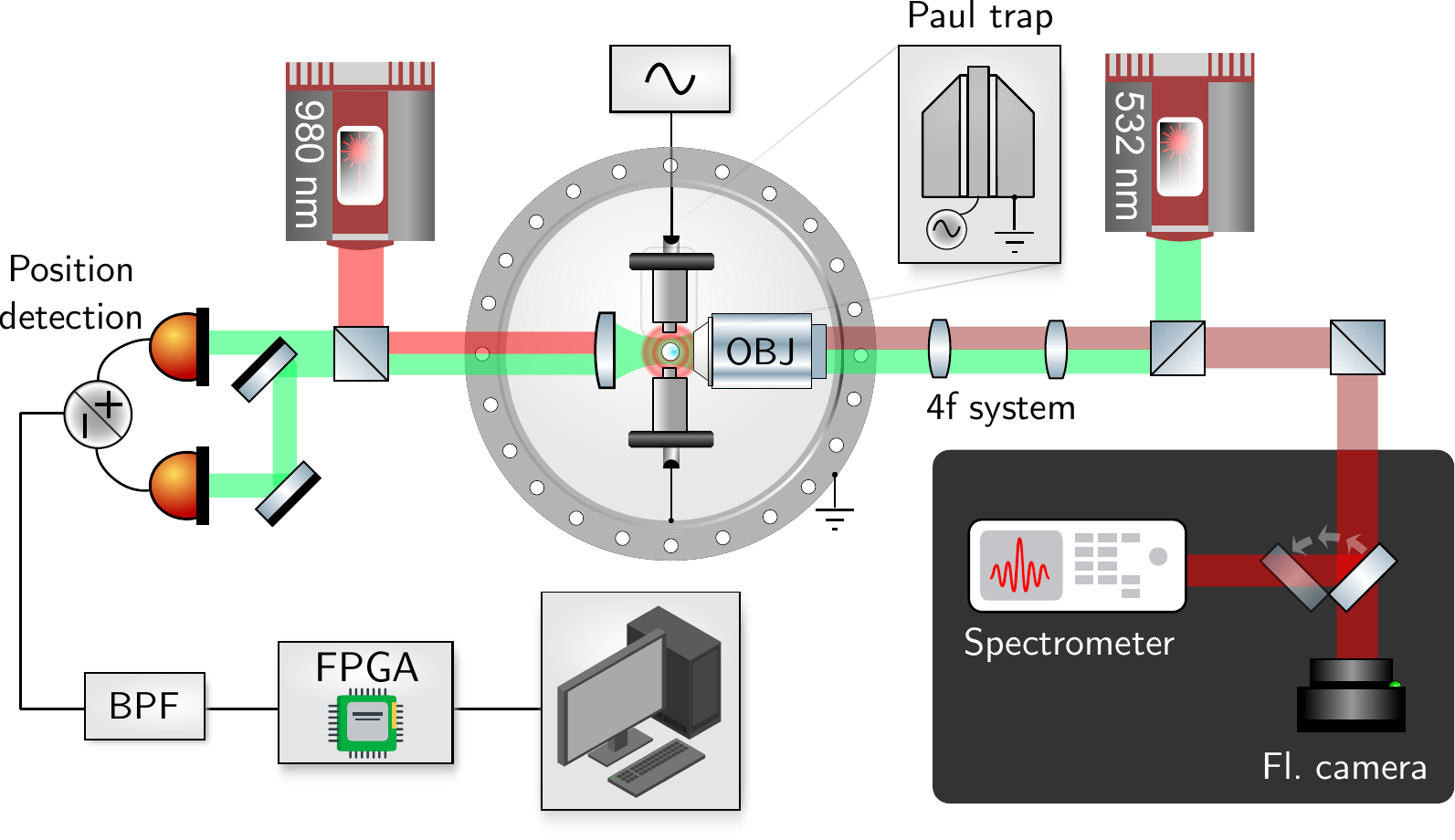}
\caption[Experimental setup]{\textbf{Experimental setup}.
A Paul trap, driven by a high voltage sinusoidal signal, levitates a charged nanoparticle. The particle is illuminated from the left with a weakly focused 980 nm diode laser to facilitate the loading process. Once the particle is in the trap, the 980~nm beam is switched off and a 532~nm laser focalized with a 0.8 NA objective (OBJ) is turned on, adding the optical potential to the hybrid trap. The scattered light is collected with an aspheric lens and sent to a quadrant photodiode for motion detection. The particle signal is band-pass filtered (BPF) and sent to an FPGA and computer, where it is processed. The 532~nm laser can also excite internal transitions, in which case the fluorescence is collected and measured in a light-tight box.}
\label{fig:S1}
\end{center}
\end{figure} 
 
\phantomsection
\subsection{Numerical simulations}\label{sec:numerical}

We used the code from Ref.~\citep{conangla2020overdamped}, containing functions and libraries in C++ to generate sample paths of a vector process (of arbitrary dimension) defined by a stochastic differential equation
\begin{align}
\dif \textbf{X} = \textbf{a}(t, \textbf{X})\,\dif t+ \textbf{b}(t, \textbf{X})\,\dif W.
\end{align}
The simulation of $X_t$ is performed with a modified Runge-Kutta method for stochastic differential equations~\citep{roberts2012modify} (strong order 1, deterministic order 2), detailed at the end of the section. This particular method does not require any non-zero derivatives of the diffusion term $\textbf{b}(t, \textbf{X})$. Other methods (e.g. the Milstein method) have strong order 1 but reduce to the Euler-Maruyama method (strong order 0.5) when $\textbf{b}(t, \textbf{X})$ is a constant.

Since the realizations of the process have a certain degree of randomness, each of them will be different and many traces (usually around $n = 1000$) need to be generated to estimate the process statistical moments 
\begin{align}
\mathbb{E}[f(X_t)] \simeq \frac{1}{n}\sum_i f(X^i_t).
\end{align}
This is usually quite intensive in terms of processing power and computer memory. For this reason, the main computation is coded in C++, while MATLAB and Python are used for post-processing. The code and libraries can be freely downloaded from \href{https://github.com/gerardpc/sde\_simulator}{https://github.com/gerardpc/sde\_simulator}. 

\phantomsection
\subsubsection{Simulation parameters}\label{sec:parameters}
For our simulations we use the following set of parameters, based on the experimental setup:
\begin{itemize}
\item $T = 295$ K (ambient temperature).
\item Particle radius: we assume $100 \leq r \leq 1000$ nm. From the radius, $m = \frac{4}{3}r^3\cdot \rho$, where $\rho$ is the material density. Assuming silica, this results in $m  \in [10^{-17} \, 10^{-15}]$ kg.
\item $m\Gamma = \gamma = 6\pi \nu r$, where $\nu = 18.6\cdot 10^{-6}$ Pa$\cdot$s is the viscosity of air at ambient pressure. This gives $\gamma = 3.5 \cdot 10^{-11}$ kg/s. $\sigma = \sqrt{2k_\text{B} T m\Gamma}$ is obtained from the fluctuation-dissipation theorem, $\sigma = 5.3 \cdot 10^{-16}$.
\item $\omega_\text{d}/2\pi \in [0.5,\, 20]$ kHz
\item $Q$ is the net number of charges in the particle (assume $50 \leq Q \leq 1000$), $V$ the electric potential at the electrodes ($500 \leq V \leq 2000$ volts) and $d^2$ a constant factor that takes into account the geometry of the trap ($0.1 < d < 1$ mm). From these we calculate $\epsilon = \frac{Q V}{d^2}$. $\epsilon \in [4\cdot 10^{-9} \quad 3\cdot 10^{-5}]$
\end{itemize}
\vphantom{0.5cm}

\phantomsection
\subsection{Gold nanorods}\label{sec:nanorods}

Although the particles are not exactly rods, the dimensions of an equivalent rod volume are given for indicative purposes: $33\text{ nm}\times63\text{ nm}$. The gold nanorods were PEGylated with HS-PEG-OMe (2000Da) and re-dispersed in MQ water. Its surface plasmon resonance can be found in Fig. \ref{fig:5}.

\phantomsection
\subsection{Dielectric nanoparticles}\label{sec:nanoparticles}

References for the nanoparticles used:
\begin{enumerate}
\item \textbf{Silica particles} $d = 143$ nm monodisperse from micro particles GmbH.
\item \textbf{Nanodiamonds} $d = 70$ nm from Adámas Nano.
\item \textbf{Er+ doped particles} selected with a 200 nm porous filter.
\end{enumerate}


\end{document}